\documentclass[12pt]{article}
\usepackage{geometry}                		% See geometry.pdf to learn the layout options. There are lots.
\usepackage{graphicx}				% Use pdf, png, jpg, or eps§ with pdflatex; use eps in DVI mode
\usepackage{amsmath}								% TeX will automatically convert eps --> pdf in pdflatex		
\usepackage{amssymb}
\usepackage{amscd}
\usepackage{yfonts}
\usepackage{amsmath,amsthm,amssymb}
\usepackage{amsthm}
\newtheorem{theorem}{Theorem}

              \newtheorem{lemma}[theorem]{Lemma}

\newcommand {\bp} {\bf p}
\newcommand {{\bx}} {\bf x}
\newcommand {{\bk}} {\bf k}
\newcommand {\bP} {\bf  P}

\date{}

\medskip

\title{Scattering in  algebraic  approach to quantum theory. Jordan algebras}
\author{Albert Schwarz}
\date{}							% Activate to display a given date or no date

\begin{document}
\author {  A. Schwarz\\ Department of Mathematics\\ 
University of 
California \\ Davis, CA 95616, USA,\\ schwarz @math.ucdavis.edu}

%%Albert Schwarz, \textsc{Department of Mathematics, University of California, Davis} \par \nopagebreak \textit{E-mail address:} \texttt{schwarz@math.ucdavis.edu}
%====}}

\maketitle

%\begin{document}

 \begin{abstract}
  Using geometric approach  we formulate quantum theory in terms of Jordan algebras. We analyze the notion of  (quasi)particle (=elementary excitation of translation-invariant stationary state) and the scattering of  (quasi)particles in this framework. 
 \end{abstract}
  
\section {Introduction}
In algebraic approach physical observables correspond to self-adjoint elements  of $*$-algebra $\cal A.$  The vector space $\cal B$  of self-adjoint elements of $\cal A$ is not closed with respect to the operation of multiplication, but it is closed with respect to the operation $a\circ b=\frac 1 2 (ab+ba).$  It was suggested long ago \cite {J} that more natural algebraic approach should be based on 
axiomatization of the operation $a\circ b.$ This idea led to the notion of Jordan algebra defined as a commutative algebra  over $\mathbb{R}$ with multiplication $a\circ b$ obeying   \begin{equation}\label {ID}(x\circ y)(x\circ x)= x\circ (y\circ (x\circ x)).\end{equation}(Defining by $R_a$ the operator of multiplication by $a$ we can express this identity saying that operators $R_a$ and $R_{a\circ a}$ commute.)

The space $\cal B$ is closed also with respect to the linear operator $Q_a$ transforming $x\in \cal B$ into $axa$ (here $a\in\cal B$). Another approach to Jordan algebras is based on axiomatization of this operator. The operator $Q_a$ is quadratic with respect to $a$, hence it can be extended to  operator $Q_{\tilde a,a}$ that is symmetric and bilinear with respect to $\tilde a, a.$ Imposing some conditions on $Q_a$ we obtain the notion of quadratic Jordan algebra.
Starting with original definition of Jordan algebra we obtain a quadratic Jordan algebra taking $Q_a= 2R(a)^2-R(a\circ a). $
Conversely starting with quadratic Jordan algebra $\cal B$  we obtain a family of products obeying (\ref{ID}) by the formula

$$a\circ_x b=Q_{a,b}(x).$$
%If $x\in \cal B$ is invertible (I.e.$R_x$ is invertible) such a product specifies a unital Jordan algebra.
In what follows we  work with unital topological Jordan algebra $\cal B$ specified by a product $a\circ b$. It seems, however, that
quadratic Jordan algebras are more convenient to relate Jordan algebras to geometric approach \cite{SG}-\cite{GA3}. This relation is
based on the remark that   one can define a cone ${\cal B}_+$ in $\cal B$ as the smallest convex closed set , invariant with respect to operators $Q_a$ and containing the unit element. One can consider also the dual cone
${\cal B}_+^{\vee}$ consisting of linear functionals on $\cal B$ that are non-negative on ${\cal B}_+.$ 
( See Section 2 for more details).

In the case when $\cal B$ consists of self-adjoint elements of $C^*$-algebra $\cal A$ elements of ${\cal B}_+^{\vee}$
can be identified with positive linear functionals  on $\cal A$ (states). This means that applying geometric approach to Jordan algebras we generalize the algebraic approach based on $*$-algebras. 
Notice, that  Jordan algebras  can be regarded as the natural framework of algebraic approach. This statement is prompted  by the following
 theorem: Cones of states of two $C^*$-algebras are isomorphic iff corresponding Jordan algebras are isomorphic ( Alfsen - Shultz).

One says that $\cal B$ is a JB-algebra if it is equipped with Banach norm obeying 
$$ ||x\circ y||\leq ||x||\cdot ||y||, ||x^2||=||x||^2, ||x^2||\leq ||x^2+y^2||.$$ 
For such an algebra the cone ${\cal B}_+$ consists of squares. ( For any Jordan algebra an element
$a\circ a=Q_a(1)$ belongs to the cone; for JB-algebras all elements of the cone have this form.) It follows that 
for JB-algebras the cones are homogeneous:  automorphism groups of cones act transitively on the interior of the cone.  

It is natural  to use homogeneous cones in geometric approach, hence it seems that JB-algebras can lead to interesting models.

The appearance of cones in the theory of Jordan algebras allows us to apply  general constructions of geometric approach to these algebras. In the Section 3 we consider (quasi)particles in the framework of Jordan algebras. In Sections 4 and 5 we consider scattering of  (quasi)particles. In Section 6 we define generalized Green functions and show that (inclusive) scattering matrix can be expressed in terms of these functions. Section 7 is devoted to some generalizations of Jordan algebras and of above results.

We do not discuss here numerous papers where the theory of Jordan algebras is related to physics
(see, in particular, \cite {FJGAU}-\cite{IORD}).  A short review of the theory of Jordan algebras and Jordan pairs as well as a review of various relations between Jordan algebras and physics is given in a companion paper \cite {GA44}.

\section {Jordan algebras}

Let us consider a topological Jordan algebra over  $\mathbb {R}$  denoted $\cal B.$ (Recall that Jordan algebra is defined as a  commutative algebra with multiplication obeying the identity $(x\circ y)(x\circ x)= x\circ (y\circ (x\circ x)).$ In what follows we consider unital Jordan algebras.  If $\cal B$ is a complete topological vector space and the multiplication is continuous we say that $\cal B$ is a topological Jordan algebra.) The most important class of topological Jordan algebras consists of Jordan Banach algebras ($JB$-algebras). The Banach norm in JB-algebra should obey
 $$ ||x\circ y||\leq ||x||\cdot ||y||, ||x^2||=||x||^2, ||x^2||\leq ||x^2+y^2||.$$ 
  (See \cite {HO} for a review of the theory of operator Jordan algebras.)
  
  In finite-dimensional case the class of JB-algebras coincides with the class of  Euclidean Jordan algebras classified in the famous paper by Jordan,  von Neumann and Wigner \cite {JNW}.  The most natural simple finite-dimensional JB-algebras $\textgoth {h}_n (\mathbb {R}), \textgoth {h}_n (\mathbb{ C}), \textgoth {h}_n (\mathbb{H})$  consist of Hermitian matrices  with real, complex or quaternion entries. One more series of simple finite-dimensional JB-algebras is the series of spin factors  (or Jordan algebras of Clifford type). These algebras are  generated by elements $1,e_1,..,e_n$ with relations $e_i\circ e_i=1, e_i\circ e_j=0$ for $i\neq j$ .
The last simple finite dimensional JB-algebra is 
Albert algebra $\textgoth {h}_3 (\mathbb {O})$ that can be realized as an algebra of  $3\times 3$ Hermitian matrices with octonion entries. This  $27$-dimensional algebra is exceptional ( it cannot be embedded into matrix Jordan algebra with the operation $a\circ b=\frac 1 2 (ab+ba).$).

The structure semigroup $Str (\cal B)$ is  defined as a semigroup generated
by automorphisms of $\cal B$ and operators $Q_a$ that can be expressed in terms of Jordan triple product $\{a,x,b\}$ by the formula $Q_a(x)=\{a,x,a\}.$ (Jordan triple product is defined by the formula
$\{a,x,b\}=(a\circ x)\circ b+(x\circ b)\circ a-(a\circ b)\circ x.)$
The operator $Q_a$ is quadratic with repect to $a$; we use the notation $Q_{\tilde a, a}$ for the corresponding bilinear operator: $Q_{\tilde a, a}=\{\tilde a,x,a\}.$

An element $B\in Str (\cal B)$   ( a structural transformation) obeys
\begin {equation}
\label {QQ}
Q_{Ba}= BQ_aB^t
\end {equation}
where $B\to B^t$ denotes  an involution in the structure semigroup transforming every operator $Q_a$ into itself and every automorphism into inverse automorphism. 
The structure group $Strg (\cal B)$ is generated by automorphisms of $\cal B$ and invertible operators $Q_a$. 

The inner structure semigroup $iStr(\cal B)$ is generated by operators $Q_a$. The inner structure group
$iStrg(\cal B)$ consists of invertible  elements of  inner structure semigroup.

We define the positive cone in Jordan algebra  as the smallest closed convex subset of $\cal B$ containing the unit element and invariant with respect to operators $Q_a$.\footnote { We  define a cone in  topological vector space as a closed convex subset, that is invariant with respect to dilations $x\to \lambda x$ where $\lambda>0.$ We do not impose any further restrictions, hence  in our terminology   a vector space is a cone.} One can say also that the positive cone is the smallest closed convex subset of $\cal B$ that contains unit element and is invariant with respect to inner structure semigroup.   Positive cone is invariant with respect to  automorphisms, hence it is invariant also with respect to  the structure semigroup. It is obvious that  the cone  contains all squares; this follows from $Q_a(1)=a\circ a.$ For JB-algebras all elements of the cone can be represented as squares and the structure group acts transitively on the interior of the cone. 

The positive cone is denoted by ${\cal B}_+$ and the dual cone is denoted by ${\cal B}_+^{\vee}.$   % The structure semigroup acts on both  cones.

Let us suppose that the Jordan algebra $\cal B$ is obtained from associative algebra $\cal A$ as a set of  self-adjoint elements with respect to involution $^*$; this set is equipped  with the operation $a\circ b=\frac {1}{2}(ab+ba).$ Then  $Q_a(x)=axa, Q_{\tilde a, a}=\frac {1}{2}(\tilde a xa+ax\tilde a).$ It follows that the structure semigroup of $\cal B$ contains  all maps $x\to A^*xA$ where $A$ is a self-adjoint or unitary element of $\cal A$ (for self-adjoint element we get a map $Q_A$, for unitary element we get an automorphism).

 If $\cal  A$ is a $C^*$-algebra  then every element of the form $A^*A$ can be represented as a square of self-adjoint element, hence  the cone in $\cal B$ is a smallest closed convex set containing all elements of the form $A^*A$ where $A\in \cal A.$
The dual cone consists of all linear functionals on $\cal B$ that are non-negative on the elements of the form $A^*A$ (in agreement with the standard definition of positive linear functional on associative algebra with involution).

We consider also complexifications of  Jordan algebras considered  as complex
Jordan algebras with involution. ( If we start with  $JB$-algebra then the complexification is called $JB^*$-algebra.)The cones associated with these algebras can be defined as the cones of their real parts.
The Jordan triple product $\{a,x^*,b\}$  is defined as an operation antilinear with respect to the middle arguments and linear with respect to other arguments. 
We introduce the notation $Q_a(x)=\{a,x,a^*\}$  where $x$ is real. It is easy to check that $Q_a$ belongs to the structure semigroup (it is equal to $Q_{\alpha}+Q_{\beta}$ where $\alpha$ and $\beta$ are real and imaginary parts of $a$). It follows
that in the case when $x$ is real and $b=a^*$ the triple product belongs to the positive cone. Notice that the map $Q_a$ is Hermitian with respect to $a.$

In geometric approach to scattering theory \cite {GA3} we are starting with a cone  of states $\cal C\subset \cal L$  and a group $\cal U$ consisting of automorphisms of the cone. ( Here $\cal L$ denotes a complete topological vector space.)

   If we take as a starting point a Jordan algebra $\cal B$ we can
take as $\cal C$ either the cone ${\cal B}_+$ or the dual cone.  The group $\cal U$ can be identified with the structure group  $Strg (\cal B)$.  Sometimes it is convenient to fix a semiring $\cal W$ consisting of endomorphisms of the cone; if we are starting with Jordan algebra $\cal B$ the semiring $\cal W$ can be defined as the smallest semiring containing the structure semigroup  $Str (\cal B)$.

\section { (Quasi)particles}

To define (quasi)particles and their scattering we should specify time translations $T_{\tau}$ and spatial translations $T_{\bx}$ as elements of the structure group $Strg (\cal B)$. In other words we should fix a homomorphism of the commutative translation group $\cal T$ to $Strg (\cal B)$. The translation group acts also on the cones. We are using the same notations $T_{\tau}, T_{\bx}$  for time and spatial translations of the cones.  As usual we denote $T_{\tau} T_{\bx}\alpha$ as $\alpha (\tau,\bx).$
As in \cite {SGA}, \cite {GA3} we define (quasi)particles  as elementary excitations of translation-invariant stationary state.

Applying the  general definition of geometric approach we can say that an elementary excitation of  stationary translation- invariant state $\omega$ is a  quadratic or Hermitian  map $\sigma$ of the "elementary space" $\textgoth {h}$ into the cone of  states $\cal C$. This map should commute with spatial and time translations. In addition one should fix a map of $\textgoth {h}$ into the space $End(\cal L)$  of endomorphisms of $\cal L$ such that  $\sigma(f)= L(f)\omega$  \cite {GA3}. In the framework of Jordan algebras  $\omega$ is an element
of the positive cone ${\cal B}_+$ or of the dual cone ${\cal B}_+^{\vee}.$ For definiteness we assume that $\omega$ is an element of the dual cone; then $\cal L$ should be identified with ${\cal B}^{\vee}.$ (Recall that that the elementary space
$\textgoth {h}$ is defined as a pre Hilbert space of smooth fast decreasing functions depending on ${\bx}\in \mathbb {R}^d$ and discrete variable $i\in \cal I.$ The spatial translations act as shifts by ${\bf a}\in \mathbb {R}^d$ , the time translations commute with spatial translations. We can consider real-valued or complex-valued functions; we should consider quadratic maps in the first case and Hermitian maps in the second case.)

Let us start with linear map $\rho:\textgoth {h}\to \cal B$ commuting with translations. Let us fix translation-invariant stationary state $\omega\in {\cal B}_+^{\vee}$ obeying  additional condition $T^t\omega=\omega$ for all $T\in \cal T$. ( Here $T\to T^t$ stands for the involution in  structure group  entering (\ref {QQ}).) Then {\it the map $L:\textgoth {h}\to End ({\cal B}^{\vee})$ transforming 
$x\in \textgoth {h}$ into $Q_{\rho (x)}$  specifies an elementary excitation of $\omega$  as the map}  $\sigma: x\to L(x)\omega$. To verify this statement we notice that
the map $\sigma$ is quadratic  or Hermitian because $L$ is quadratic  (if the $\cal B$ is a Jordan algebra over $\mathbb {R}$) or Hermitian (if $\cal B$ is a complex Jordan algebra with involution), the formula (\ref {QQ}) implies that $\sigma$ commutes with translations. In what follows we consider  mostly elementary excitations  constructed this way. Notice, however,
that we can start  with arbitrary   linear map $\rho:\textgoth {h}\to \cal B$ and impose a weaker condition  that the map $\sigma: x\to L(x)\omega$  commutes with translations. Then $\sigma$ can be regarded as elementary excitation. (Here again  $L$ transforms $x\in \textgoth {h}$ into $Q_{\rho (x)}$.)

Let us consider as an example  a Jordan algebra defined as a set of of self-adjoint elements  of Weyl algebra. We define Weyl algebra corresponding to real pre Hilbert space $\textgoth {h}$ as a unital $*$- algebra generated by elements $a(f), a^+(g)$ depending linearly of $f,g\in \textgoth {h}$ and obeying canonical commutation relations
$$[a(f), a(g)]=[a^+(f),a^+(g)]=0, [a(f),a^+(g)]=\langle f,g \rangle.$$

We define  a map $\rho$ of $\textgoth {h}$ into this Jordan algebra as a map sending $f\in \textgoth {h}$ to $a(f)+a^+(f).$ 

Let us assume that  $\textgoth {h}$ is an elementary  space.  Then the translations in $\textgoth {h}$ induce  translations in  Weyl algebra and in the corresponding Jordan algebra.  The map $\rho$ 
commutes with translations. Taking as $\omega$ the state corresponding  to the Fock vacuum in positive cone or a dual cone we obtain an example of elementary excitation of this state.

The same construction works for Clifford algebra specified by canonical anticommutation relations.

Notice that  the Jordan algebra corresponding to Clifford algebra can be regarded as $ JB$-algebra, but starting with Weyl algebra we obtain a topological Jordan algebra, more precisely a Fr\'echet Jordan algebra.  (Fr\'echet vector space is a complete topological vector space where the topology is specified by a countable family of seminorms. In what follows we are talking about $JB$-algebras, but our results can be generalized to Fr\'echet analogs of these algebras.)

Let us discuss the relation of the above constructions to the construction of elementary excitations in the approach based on consideration of associative algebra $\cal A$ with involution ($*$-algebra).
The set of self-adjoint elements  of such an algebra can be regarded as Jordan algebra $\cal B$ over $\mathbb {R}$; the complexification $\mathbb{C}\cal B$ of this Jordan algebra can be identified with  $\cal A$  considered  as complex Jordan algebra with involution.  The elements of  dual cones of these Jordan algebras  can be identified with not necessarily normalized states of $\cal A.$ Let us assume that  spatial and time translations act on $\cal A$ as automorphisms; this action generates an action of translations on the cones of Jordan algebras. Let us fix a translation invariant stationary state $\omega$ of algebra $\cal A$; corresponding elements of dual cones of Jordan algebras are denoted by the same symbol.  Excitations of $\omega$ can be regarded as elements of pre Hilbert space $\cal H$ obtained by means of GNS construction applied to $\omega,$  an elementary excitation is an isometric embedding $\Phi (f)$ of elementary space $\textgoth {h}$ into $\cal H.$ Following \cite {GA3} we represent 
$\Phi(f)$ in the form $B(f)\theta $ where $B$ is a linear map $\textgoth {h}\to \cal A$ and $\theta \in \cal H$ is a vector corresponding to the state $\omega$. If elements $B(f)$ are self-adjoint  we can apply the above construction of  elementary excitation of $\omega$ considered as an element of a dual cone of the Jordan algebra $\cal B$ of self-adjoint elements of $\cal A$ taking $\rho=B$. Then the state $Q_{\rho(f)}\omega$ corresponds to the vector $B(f)\theta.$ If $B^*(f)=0$  a similar statement  can be proved for $\mathbb{C}\cal B$ ( for Jordan algebra  with involution  obtained by  complexification of $\cal B$).
\section {Scattering}
To analyze scattering in the framework of Jordan algebras we are starting with  linear map $\rho:\textgoth {h}\to \cal B$ (not necessarily  commuting with translations). We fix translation-invariant stationary state $\omega\in {\cal B}_+^{\vee}$.  The map $L:\textgoth {h}\to End ({\cal B}^{\vee})$ transforming 
$x\in \textgoth {h}$ into $Q_{\rho (x)}$  specifies an elementary excitation of $\omega$  if  the map  $\sigma: x\to L(x)\omega$ commutes with translations. 

We define ( following the general theory of \cite {GA3}) the operator
$$L(f,\tau)= T_{\tau}( L(T_{-\tau}f))= T_{\tau}L(T_{-\tau}f)T_{-\tau}.$$
 where $f\in \textgoth {h}.$  This operator is quadratic or Hermitian with respect to $f$, therefore we can consider also the  operator $L(\tilde f,f,\tau)$ that is linear with respect to $\tilde f$ and linear or antilinear with respect to $f$;  it coincides with $L(f,\tau)$ for $\tilde f= f^*.$ ( Notice that in the case of real vector spaces $f^*=f.$) Using the notation $Q_{\tilde a,a}x=\{\tilde a,x,a^*\}$ we can write $$L(\tilde g,g,\tau)=T_{\tau} Q_{T_{-\tau}\rho (\tilde g),T_{-\tau}\rho(g)}T_{-\tau}=L(\tilde g,g)T^t_{-\tau}$$
where $L(\tilde g, g)=Q_{\rho(\tilde g), \rho(g)}$ is a bilinear (or sesquilinear)  form corresponding to the  quadratic (or Hermitian) form $L(g) =Q_{\rho (g)}.$
 
 The state
 \begin {equation} \label {EQ}
 \Lambda (f_1,\cdots,f_n|-\infty)= \lim _{\tau_1\to-\infty,\cdots, \tau_n\to-\infty}
 L(f_1,\tau _1),...L(f_n,\tau_n)\omega
 \end {equation}
describes the collision of (quasi)particles with wave functions $f_1,...,f_n.$ We say that (\ref {EQ}) is a scattering state (or, more precisely, an $in$-state).

The result of the collision can be characterized by the number
\begin {equation}\label {BBKK}
\lim _{\tau'\to+\infty,\tau\to -\infty}\langle \alpha|L(g_1,\tau')...L(g_m, \tau') L(f_1,\tau)...L(f_n, \tau)|\omega\rangle
\end{equation}
where $\alpha$ is a stationary translation-invariant point of the  cone  ${\cal B}_+$ or of the larger cone ${\cal B}_+^{\vee \vee}$ (we use bra-ket notations). Comparing with the formulas of \cite {GA3} we see that this number can be interpreted as a generalization of inclusive scattering matrix. More generally we can consider
a functional
$$\sigma (\tilde g'_1,g'_1,..., \tilde g'_{n'}, g'_{n'},\tilde g_1,g_1,...,\tilde g_n, g_n)=$$
\begin {equation} \label {BKKB}
\langle \alpha |  \lim _{\tau'_i\to +\infty, \tau_j\to -\infty} L(\tilde g'_1,g'_1,\tau' _1)...L(\tilde g'_{n'}, g'_{n'},\tau'_{n'})L(\tilde g_1,g_1,\tau _1)...L(\tilde g_n,g_n,\tau_n)|\omega\rangle
\end {equation}
that is linear or antilinear with respect to all of its arguments. It also can be regarded as inclusive scattering matrix.

It was proven in \cite {GA3} that the limit (\ref {EQ})
exists  for    $f_1, \cdots, f_n$ in a dense open subset    of $\textgoth {h}\times \cdots \times \textgoth {h} $  if
\begin{equation}\label{COM}
||[T_{\alpha}(L(\phi)),L(\psi)]||\leq \int d{\bx} d {\bx}' D^{ab}(\bx-\bx')|\phi_{\it a}(\bx)|\cdot |\psi_{\it b}(\bx')|
\end{equation}
{\it where $D^{ab}(\bx)$ tends to zero faster than any power as $\bx\to \infty$ and $\alpha$ runs over a finite interval.}

Less formally we can formulate this condition as the requirement that the commutator 
$ [T_{\alpha}(L(\phi)),L(\psi)]$ is small if the essential supports of functions $\phi$ and $\psi$ in coordinate representation are far away.

Similar conditions can be formulated for the existence of limits  (\ref {BBKK}), (\ref {BKKB}) (for the existence of inclusive scattering matrix).
Later we will formulate more concrete conditions for the existence of the above limits.

 Notice that the linear map $\rho:\textgoth {h}\to \cal B$  can be regarded as a multicomponent generalized  function  $\rho (\bx)$ in coordinate representation or $\rho (\bk)$ in momentum representation. (This means that we formally represent $\rho(\phi)$ as
 $\int d{\bx}\phi ({\bx})\rho (\bx)$ or as $\int d{\bk}\phi ({\bk})\rho(\bk).$ Discrete indices are omitted in these formulas and  in what follows.)  We assume that the generalized functions $\rho (\bx), \rho(\bk)$ correspond to continuous functions denoted by the same symbols.
 
  The expression (\ref {BKKB}) is linear with respect to its arguments, therefore it can be regarded  as a generalized function
 
 $$\sigma (\tilde {\bx}'_1,{\bx}'_1,..., \tilde {\bx}'_{n'}, {\bx}'_{n'},\tilde {\bx}_1,{\bx}_1,...,\tilde {\bx}_n, {\bx}_n)=$$
\begin {equation} \label {BKKBB}
 \lim _{\tau'_i\to +\infty, \tau_j\to -\infty}\langle \alpha |  L(\tilde {\bx}'_1,{\bx}'_1,\tau' _1)...L(\tilde {\bx}'_{n'}, {\bx}'_{n'},\tau'_{n'})L(\tilde {\bx}_1,{\bx}_1,\tau _1)...L(\tilde {\bx}_n,{\bx}_n,\tau_n)|\omega\rangle
\end {equation}
or, in momentum representation,
 $$\sigma (\tilde {\bk}'_1,{\bk}'_1,..., \tilde {\bk}'_{n'}, {\bk}'_{n'},\tilde {\bk}_1,{\bk}_1,...,\tilde {\bk}_n, {\bk}_n)=$$
\begin {equation} \label {BKKBBB}
  \lim _{\tau'_i\to +\infty, \tau_j\to -\infty} \langle \alpha |L(\tilde {\bk}'_1,{\bk}'_1,\tau' _1)...L(\tilde {\bk}'_{n'}, {\bk}'_{n'},\tau'_{n'})L(\tilde {\bk}_1,{\bk}_1,\tau _1)...L(\tilde {\bk}_n,{\bk}_n,\tau_n)|\omega\rangle
\end {equation}.

We use the notations
$$L(\tilde g,g,\tau)=\int d \tilde {\bx}d{\bx}\tilde g(\tilde {\bx})g({\bx})L(\tilde {\bx},{\bx},\tau)=\int d \tilde {\bk}d{\bk}\tilde g(\tilde {\bk})g({\bk})L(\tilde {\bk},{\bk},\tau)$$

If $\rho$ commutes with translations we can say that
 $$T_{\tau}\rho(\phi)= \rho(T_{\tau}\phi)=\int d{\bk}e^{-i\tau E({\bk})}\rho ({\bk})\phi (\bk)
 %=\int d{\bk} \sum_je^{i\epsilon_j({\bk}) \tau}\langle a_j({\bk})\phi({\bk}),\rho ({\bk})\rangle,
$$
hence
$$L(\tilde {\bk},{\bk},\tau)=T_{\tau}(e^{-i\tau[E(\tilde{\bk})+E({\bk})]} Q_{\rho(\tilde {\bk}),\rho ({\bk})})T_{-\tau}=$$
$$e^{-i\tau[E(\tilde{\bk})+E({\bk})]}T_{\tau} Q_{\rho(\tilde {\bk}),\rho ({\bk})}T_{-\tau}$$
%$$L(\tilde g,g,\tau)=T_{\tau} Q_{T_{-\tau}\rho (\tilde g),T_{-\tau}\rho(g)}T_{-\tau}= \int d\tilde{\bk}d{\bk} L(\tilde {\bk},{\bk},\tau)$$
%where
%$$L(\tilde {\bk},{\bk},\tau)=T_{\tau}(\sum _{\tilde j,j}e^{-i[\epsilon_{\tilde j}(\tilde {\bk})+\epsilon_j
%{(\bk})]\tau} Q_{\rho ({\tilde \bk})^ta_{\tilde j}(\tilde{\bk}), \rho ({\bk})^ta_j({\bk})))}T_{-\tau}=$$
 %\begin{equation} \label {EE}\sum _{\tilde j,j}e^{-i[\epsilon_{\tilde j}(\tilde {\bk})+\epsilon_j
%({\bk})]\tau} Q_{(\rho ({\tilde \bk})^t a_{\tilde j}(\tilde{\bk}),\tau), ( \rho ({\bk})^t a_j({\bk}),\tau))}
%\end {equation}

%where the superscript $t$ stands for the matrix transposition.
 \section {Existence of inclusive scattering matrix}
 Let  us consider  first of all the case when the Jordan algebra $\cal B$ is obtained as a set of self-adjoint elements of associative Banach algebra $\cal A$ with respect to the involution $^*.$ We impose the condition of asymptotic  commutativity or anticommutativity on the multicomponent function $\rho (\bx)$:
 \begin{equation}\label {BF}
|| [\rho ({\bf x+a}),\rho({\bf a})]_{\mp}||<\frac {C_n}{1+||{\bf x}||^n}
 \end{equation}
 for every natural number $n.$
 It follows from this condition that 
 
 \begin{equation} \label {BBF}
 ||[\rho(\phi),\rho(\psi)]_{\mp}||< \int d{\bx} d {\bx}' D^{ab}(\bx-\bx')|\phi_{\it a}(\bx)|\cdot |\psi_{\it b}(\bx')|
 \end{equation}
 where $D^{ab}(\bx)$ tends to zero faster than any power as $\bx\to \infty.$

 The operators $L_{\phi}=Q_{\rho (\phi)}$ transform $ b\in \cal B$ into $\rho(\phi)b\rho(\phi).$
 It is easy to check that these operators obey (\ref{COM}), hence the limits we are interested in exist and we can consider the scattering of particles. In the case of asymptotic commutativity (anticommutativity) we are dealing with bosons (fermions).

 For Jordan algebra $\cal B$ coming from associative algebra $\cal A$  with involution it is easy to formulate 
 sufficient conditions for commutativity of operators $Q_a$ and $Q_b$.   It is obvious that
 in the case when $a$ and $b$ commute in $\cal A$ or $a$ and $b$ anticommute in $\cal A$
 (equivalently $a\circ b=0$ in $\cal B$) we have $Q_aQ_b=Q_bQ_a.$  Similar statements are
 correct in any $JB$-algebra $\cal B$:{\it  if operators $R_a$ and $R_b$ commute or $a\circ b=0$ then the operators $Q_a,Q_b$ commute} \cite {WET}, \cite {AOC}, \cite {SHE}. (Here $a,b\in \cal B$, $R_a$ stands for the operator of multiplication by $a$ in $\cal B$. If $R_a$ and $R_b$ commute one says that $a$ and $b$ operator commute.)
 
 It is natural to conjecture  that these statements can be generalized in the following way:
 
  If  $a\circ b$ is small , then the operators $Q_a$ and $Q_b$ almost commute.
  
  If   the operators $R_a$ and $R_b$ almost commute, then the operators $Q_a$ and $Q_b$ almost commute.

The first of these conjectures is proven in \cite {GA5}. More precisely,

{\it If the norm of the Jordan product $a\circ b$ of two elements of  JB-algebra $\cal B$ is $\leq \epsilon$ then
\begin {equation}
\label{E}
||[Q_a,Q_b]||\leq  k( ||a||, ||b||) \sqrt\epsilon.
\end {equation}
Here $\epsilon \geq 0$ and $k$ is a polynomial function  }

This statement allows us to give conditions for the existence of limits (\ref {EQ}), (\ref {BBKK}), (\ref {BKKB}).

We impose the condition  
\begin{equation}\label {BFF}
|| \rho ({\bf x+a})\circ \rho({\bf a})||<\frac {C_n}{1+||{\bf x}||^n}
 \end{equation}
 for every natural number $n.$
 Then
 
 \begin{equation} \label {BBF}
 ||\rho(\phi)\circ \rho(\psi)||< \int d{\bx} d {\bx}' D^{ab}(\bx-\bx')|\phi_{\it a}(\bx)|\cdot |\psi_{\it b}(\bx')|
 \end{equation}
 where $D^{ab}(\bx)$ tends to zero faster than any power as $\bx\to \infty.$
 Applying  (\ref {E}) we obtain (\ref {COM}) that implies the existence of limits (\ref {EQ}), (\ref {BBKK}), (\ref {BKKB}) for dense sets of families of functions in the arguments of these expressions.
 
 It is not clear whether the second conjecture is true. However,  the identity $Q_a=2R_a^2-R_{a^2}$  immediately implies the following weaker statement: if $a$ and $a^2$ almost operator commute with $b$ and $b^2$ then $Q_a$ and $Q_b$ almost commute:
 $$||[Q_a,Q_b]||\leq 8||a||\cdot||b||\cdot ||[R_a,R_b]||+4||a||\cdot ||[R_a,R_{b^2}||+4||b||\cdot||[R_b,R_{a^2}]||+||[R_{a^2}, R_{b^2}]|| .$$
 Using the identity $Q_{e^a}=e^{2R_a}$ one can conclude that operators $Q_{e^a}$ and $Q_{e^b}$ almost 
 commute if $a$ and $b$ almost operator commute.
 
 One can use these statements to give conditions for the existence of scattering states and
 inclusive scattering matrix.
 
 \section {Green functions}
 Let us fix translation-invariant elements $\alpha\in {\cal B}_+$, $\omega \in {\cal B}_+^{\vee}$,
 where ${\cal B}_+$ is a positive cone in  Jordan algebra $\cal B$ and ${\cal B}_+^{\vee}$ denotes the dual cone. The quadratic operators $Q_A$ act in both cones, the bilinear  operators $Q_{\tilde A,A}$ act in $\cal B$ and in the dual space $\cal B^{\vee}.$
 (Here $A,\tilde A$ are elements of $\cal B.$)
 
 Let us fix elements $\tilde A_1,A_1,..., \tilde A_n, A_n\in \cal B.$ We introduce the 
 notation $Q_i(\tilde {\bx},\tilde\tau, {\bx},\tau)$ for $ Q_{\tilde A_i(\tilde {\bx},\tilde{\tau}), A_i({\bx},\tau)}.$
 
 We define (generalized)  Green functions by the formula
 $$G_n(
 \tilde{\bx}_1,\tilde \tau_1,{\bx}_1,\tau_1,..., \tilde{\bx}_n,\tilde \tau_n,{\bx}_n,\tau_n)=$$
  $$ \langle \alpha| T( Q_1(\tilde {\bx}_1,\tilde\tau_1, {\bx}_1,\tau_1)... Q_n(\tilde {\bx}_n,\tilde\tau_n, {\bx}_n,\tau_n)| \omega \rangle $$
 where $T$ stands for the chronological ordering with respect to $\tau'_i=\frac {1}{2}(\tilde \tau_i +\tau_i).$
 
 Omitting chronological ordering in this formula we obtain a definition of correlation functions.
 Notice that in the case when Jordan algebra $\cal B$ is constructed as a set of self-adjoint elements of $*$-algebra $\cal A$ we have
 $$Q(\tilde {\bx},\tilde \tau, {\bx},\tau)a=\frac {1}{2} (A({\bx},\tau)a\tilde A(\tilde{\bx},\tilde \tau)+
\tilde A(\tilde{\bx},\tilde \tau)a A({\bx},\tau)).
 $$
 Using this remark we can express correlations functions for Jordan algebra $\cal B$ in terms of correlation functions for $*$- algebra $\cal A.$ %The situation with Green functions is more complicated, 
 
 We defined  GGreen functions in $({\bx},\tau)$-representation; as always taking Fourier transforms we obtain GGreen functions in $({\bk}, \tau)$- and $({\bk},\varepsilon)$-representations.
 
 One can show  ( under some conditions) that the inclusive scattering matrix can be calculated in terms of asymptotic behavior of GGreen functions in $({\bk}, \tau)$-representation or in terms of poles and residues in $({\bk}, \varepsilon)$-representation. The proof is similar to the proof of analogous statement in \cite {GA2}. It is based on formula (\ref{BKKBBB}).%,(\ref {EE}).
 \section {Generalizations}
 
 Let us consider a $\mathbb{Z}_2$-graded  algebra $\cal A.$ We denote by ${\cal A}_{\Lambda}$ the set of even elements of tensor product ${\cal A}\otimes \Lambda$  where 
 $\Lambda$ is a Grassmann algebra. This set can be considered as an algebra; if for every $\Lambda$  the algebra ${\cal A}_{\Lambda}$ is a Jordan algebra  we say that
 $\cal A$ is a Jordan superalgebra. (See, for example, \cite {BAR} for more standard definition of Jordan superalgebra and for  main facts of the theory of Jordan superalgebras.) 
  Similarly, if   the algebra ${\cal A}_{\Lambda}$ is a Lie algebra one says that $\cal A$ is a Lie superalgebra. One can say that  a Jordan superalgebra $\cal A$ specifies a functor defined on the category of Grassmann algebras and taking values in the category of Jordan algebras. Analogously, a Lie superalgebra specifies a functor with values in Lie algebras and a supergroup can be regarded as a functor with values in groups (all functors are defined on Grassmann algebras.
 
 Starting with a differential algebra (= $\mathbb{Z}_2$-graded  algebra equipped with an od derivation $d$ obeying $d^2=0$) we can define a differential  Jordan  superalgebra. The simplest way to construct a differential Jordan superalgebra is to take tensor product of Jordan algebra $\cal B$  by a differential supercommutative algebra $\cal E$
 (for example, one can take as $\cal E$ a free Grassmann algebra with the differential that   calculates cohomology of Lie algebra).
 
 Using these definitions one can  generalize the statements above to Jordan superalgebras and (in the framework of BRST-formalism) to differential Jordan superalgebras.
 
 These generalizations can be used to analyze interesting examples.
 
 In particular, one can construct Poincar\'e invariant theories starting with simple exceptional Jordan algebra ( Albert algebra $\textgoth {h}_3 (\mathbb {O})$). Namely, one should notice that the structure group of this algebra contains a subgroup isomorphic to  $SO(1,9)$ (this subgroup was used in \cite {FJGAU}, \cite {FJ}). Assuming that translations act trivially we   obtain  an action of ten-dimensional Poincar\'e group on
 Albert algebra.  Taking a tensor product of Albert algebra and (diffferental) supecommutative algebra where Poincar\'e group acts by automorphisms we obtain a (differential) Jordan superalgebra with action of Poincar\'e group (and in general non-trivial action of translations).


\begin{thebibliography}{10}
\bibitem {J}
Jordan, P.( 1933) Ueber die multiplikation quantenmechanischer groessen. Zeitschrift für Physik, 80(5), pp.285-291.
\bibitem {JNW} Jordan, Pascual, J. von Neumann, and Eugene P. Wigner. "On an algebraic generalization of the quantum mechanical formalism." In The Collected Works of Eugene Paul Wigner, pp. 298-333. Springer, Berlin, Heidelberg, 1993.
\bibitem {HO} Hanche-Olsen, Harald, and Erling Stormer. Jordan operator algebras. Vol. 21. Pitman Advanced Publishing Program, 1984. 


\bibitem{SG} Schwarz, A. (2020). Geometric approach to quantum theory. SIGMA. Symmetry, Integrability and Geometry: Methods and Applications, 16, 020.
\bibitem {SGA} Schwarz, A. (2021)
 Geometric and algebraic approaches to quantum theory. Nuclear Physics B, 973, p.115601.
quantum-ph 2102.09176, 
\bibitem {GA2}Schwarz, A., 2021. Scattering in algebraic approach to quantum theory. Associative algebras. arXiv preprint arXiv:2107.08553.

\bibitem {GA3}Schwarz, A., 2021. Scattering in geometric approach to quantum theory. arXiv preprint arXiv:2107



\bibitem {GA5} Schwarz, A. {Asymptotic commutativity in Jordan  algebras} (in preparation).



\bibitem {FJGAU} Foot, R. and Joshi, G.C., 1988. A natural framework for the minimal supersymmetric gauge theories. Letters in mathematical physics, 15(3), pp.237-242.

\bibitem {FJ} Foot, R. and Joshi, G.C., 1989. Space-time symmetries of superstring and Jordan algebras. International journal of theoretical physics, 28(12), pp.1449-1462.
\bibitem {GUNA} Günaydin, M., 1991. N= 2 superconformal algebras and Jordan triple systems. Physics Letters B, 255(1), pp.46-50.
\bibitem{GUNB} Günaydin, M., Sierra, G. and Townsend, P.K., 1984. The geometry of N= 2 Maxwell-Einstein supergravity and Jordan algebras. Nuclear Physics B, 242(1), pp.244-268.
\bibitem {GUNC} Günaydin, M., Sierra, G. and Townsend, P.K., 1985. Gauging the d= 5 Maxwell/Einstein supergravity theories: more on Jordan algebras. Nuclear Physics B, 253, pp.573-608.
\bibitem {CONF} Günaydin, M., 1993. Generalized conformal and superconformal group actions and Jordan algebras. Modern Physics Letters A, 8(15), pp.1407-1
\bibitem{BAEZ} Baez, J., 2002. The octonions. Bulletin of the american mathematical society, 39(2), pp.145-205.
\bibitem{BAEZA} Baez, J.C., 2012. Division algebras and quantum theory. Foundations of Physics, 42(7), pp.819-855.
\bibitem {TDV1}Todorov, I. and Dubois-Violette, M., 2018. Deducing the symmetry of the standard model from the automorphism and structure groups of the exceptional Jordan algebra. International Journal of Modern Physics A, 33(20), p.1850118.
\bibitem{TOD} Todorov, Ivan. "Exceptional quantum algebra for the standard model of particle physics." In International Workshop on Lie Theory and Its Applications in Physics, pp. 29-52. Springer, Singapore, 2019.
\bibitem {TDV2} Dubois-Violette, M. and Todorov, I., 2019. Exceptional quantum geometry and particle physics II. Nuclear Physics B, 938, pp.751-761.
\bibitem {TDV3} Dubois-Violette, M. and Todorov, I., 2020. Superconnection in the spin factor approach to particle physics. Nuclear Physics B, 957, p.115065.
\bibitem {TODOR}Todorov, Ivan, and Svetla Drenska. "Octonions, Exceptional Jordan Algebra and The Role of The Group $ F_4 $ in Particle Physics." Advances in Applied Clifford Algebras 28, no. 4 (2018): 1-36.
\bibitem{BOY} Boyle, L., 2020. The standard model, the exceptional Jordan algebra, and triality. arXiv preprint arXiv:2006.16265.

\bibitem{IORD} Iordanescu, R., 2011. Jordan structures in mathematics and physics. arXiv preprint arXiv:1106.4415.
\bibitem {GA44} Schwarz,A., Jordan algebras, Jordan pairs and physics. Review, problems, conjectures (in preparation)
%\bibitem {GA5}  Schwarz, A. Asymptotic commutativity in Jordan  algebras (in preparation)
\bibitem {AOC}Anquela, Jos\'e, Teresa Cort\'es, and Holger Petersson. "Commuting $U_a$-operators in Jordan algebras." Transactions of the American Mathematical Society 366.11 (2014): 5877-5902.=

\bibitem {SHE} Shestakov, Ivan. "On commuting U-operators in Jordan algebras." Non-Associative and Non-Commutative Algebra and Operator Theory. Springer, Cham, 2016. 105-109.

\bibitem {WET}van de Wetering, John. "Commutativity in Jordan Operator Algebras." Journal of Pure and Applied Algebra (2020): 106407.

\bibitem {BAR}
Barbier, S. and Coulembier, K., 2018. On structure and TKK algebras for Jordan superalgebras. Communications in Algebra, 46(2), pp.684-704.

\end {thebibliography}
\end {document}